\documentclass[conference]{IEEEtran}
\IEEEoverridecommandlockouts
\usepackage{cite}
\usepackage{amsmath,amssymb,amsfonts}
\usepackage{algorithmic}
\usepackage{graphicx}
\usepackage{textcomp}
\usepackage{xcolor}
\usepackage{adjustbox}
\usepackage{comment}
\usepackage{xurl}

\def\BibTeX{{\rm B\kern-.05em{\sc i\kern-.025em b}\kern-.08em
    T\kern-.1667em\lower.7ex\hbox{E}\kern-.125emX}}

\newcommand{\etal}{\textit{et al}. }

\newcommand{\eg}{\textit{e}.\textit{g}., }

\begin{document}

\title{Excitements and Concerns in the Post-ChatGPT Era: Deciphering Public Perception of AI through Social Media Analysis \\
}

\author{
\IEEEauthorblockN{Weihong Qi}
\IEEEauthorblockA{ \textit{Department of Political Science} \\
\textit{University of Rochester}\\
Rochester, USA \\
wqi3@ur.rochester.edu}

\and
\IEEEauthorblockN{Jinsheng Pan, Hanjia Lyu, Jiebo Luo}
\IEEEauthorblockA{ \textit{Department of Computer Science} \\
\textit{University of Rochester}\\
Rochester, USA \\
\{jpan24, hlyu5\}@ur.rochester.edu, jluo@cs.rochester.edu}
}
\maketitle

\begin{abstract}
As AI systems become increasingly prevalent in various aspects of daily life, gaining a comprehensive understanding of public perception towards these AI systems has become increasingly essential for several reasons such as ethical considerations, user experience, fear, disinformation, regulation, collaboration, and co-creation. In this study, we investigate how mass social media users perceive the recent rise of AI frameworks such as ChatGPT. We collect a total of 33,912 comments in 388 unique subreddits spanning from November 30, 2022 to June 8, 2023 using a list of AI-related keywords. We employ BERTopic to uncover the major themes regarding AI on Reddit. Additionally, we seek to gain deeper insights into public opinion by examining the distribution of topics across different subreddits. We observe that technology-related subreddits predominantly focus on the technical aspects of AI models. On the other hand, non-tech subreddits show greater interest in social issues such as concerns about job replacement or furlough. We leverage zero-shot prompting to analyze the sentiment and perception of AI among individual users. Through a comprehensive sentiment and emotion analysis, we discover that tech-centric communities exhibit greater polarization compared to non-tech communities when discussing AI topics. This research contributes to our broader understanding of public opinion surrounding artificial intelligence.
\end{abstract}


\section{Introduction}
Artificial Intelligence (AI) has become increasingly pervasive in our lives, transforming various sectors and shaping the future of technology. As AI continues to advance and integrate into society, it is essential to gain insights into how the general public perceives this transformative technology. Public perception plays a crucial role in the adoption, acceptance, and ethical considerations surrounding AI. The recent surge in discussions about \textit{generative AI}, exemplified by models like ChatGPT, highlights the growing interest and excitement surrounding this technology. There was a substantial surge in online discussions about AI during the month of April 2023, coinciding with the presumed release of GPT-4. The development of advanced language models has paved the way for generating human-like text, enabling realistic and interactive conversations with AI systems. ChatGPT was estimated to have reached 100 million monthly active users in January, 2023, just two months after its launch~\cite{reuters2023chatgpt}. In spite of its widespread popularity, ChatGPT has also raised concerns regarding various issues, including but not limited to data privacy. The use and adoption of ChatGPT have sparked discussions and debates surrounding potential risks associated with the handling and storage of user data. For instance, OpenAI's introduction of privacy controls played a pivotal role in Italy's decision to lift the ban on ChatGPT due to privacy concerns~\cite{bbc2023chatgpt}. Prior to OpenAI's announcement, Italy had maintained restrictions on the use of ChatGPT over apprehensions about privacy implications~\cite{bbc2023ban}.

As of May 2023, while a majority of Americans have become aware of ChatGPT, only a limited number have actually engaged with the technology themselves~\cite{pew2023majority}. The public perception of artificial intelligence is still in the process of being shaped and evolving. Therefore, it becomes crucial to comprehend and analyze public opinion surrounding AI. By understanding the viewpoints, attitudes, and concerns of the general public, we can gain valuable insights into the current state of public perception, bridge knowledge gaps, and ensure that the development and deployment of AI technologies align with societal expectations and values.

In order to gain insights into public perception of artificial intelligence, \textbf{particularly with regard to the emerging field of generative AI}, we conduct a data collection from Reddit. Our objective is to explore the thematic and sentiment attributes of online discussions surrounding this topic. We aim to answer two research questions:

\begin{itemize}
    \item RQ 1: What specific topics characterize the discussions of AI on Reddit? How do the topics vary across subreddits?
    \item RQ 2: What is the prevailing sentiment surrounding the most discussed topics, and do these sentiments differ among subreddits?
\end{itemize}

Our approach consists of BERTopic topic modeling~\cite{grootendorst2022bertopic}, zero-shot prompting for sentiment analysis~\cite{ouyang2022training}, the Linguistic Inquiry and Word Count (LIWC) text analuysis~\cite{tausczik2010psychological}, and regression analysis. By delving into these topic distributions within various subreddits, we identify differing areas of emphasis and concerns among different user communities. In order to gauge the sentiments expressed in the comments, we infer and compare sentiments both at the topic level and across different subreddits. Our findings shed light on the prominent themes, the varying concerns across different subreddits, and the sentiments expressed in relation to these topics.

\section{Related Work}

Several studies have explored public perceptions and discussions surrounding artificial intelligence, with a particular focus on generative AI and ChatGPT. Miyazaki~\etal~\cite{miyazaki2023public} investigated users' perceptions of generative AI on Twitter, especially focusing on their occupation and usage. The findings reveal that a significant interest in generative AI extends beyond IT-related occupations to encompass individuals across various professional domains. Leiter~\etal~\cite{leiter2023chatgpt} analyzed over 300,000 tweets and more than 150 scientific papers to investigate how ChatGPT is perceived and discussed. The general consensus regarding ChatGPT is that it is perceived as a high-quality system, with positive sentiment prevailing and emotions of joy dominating social media discussions. Furthermore, recent scientific papers portray ChatGPT as a promising opportunity in diverse fields, including the medical domain. However, ethical concerns surrounding ChatGPT's capabilities are also acknowledged, highlighting its potential as a double-edged sword. In the context of education, assessments of ChatGPT are mixed, with varying opinions on its impact and efficacy. These findings are align with Tlili~\etal~\cite{tlili2023if} and Sullivian~\etal~\cite{sullivan2023chatgpt}.

While other social platforms have their own unique advantages and characteristics, Reddit's combination of diverse communities, long-form discussions, user anonymity, and data availability make it a valuable source for conducting text analysis and gaining deeper insights into public opinions and discussions. More specifically, Reddit hosts a vast array of communities called subreddits, each focused on specific topics or interests. This diversity allows researchers to analyze discussions within dedicated communities, providing more focused and specialized insights. In addition, unlike platforms that primarily rely on short-form content like tweets, Reddit facilitates in-depth discussions with longer posts and comments. This allows for more detailed and nuanced analysis of user opinions, arguments, and perspectives~\cite{wu2021characterizing, waller2021quantifying,zhou2022fine}. Hence, we choose to conduct our study using Reddit data, leveraging its unique attributes to gain deeper insights into the multifaceted landscape of AI discussions.

\section{Method}\label{AA}

\subsection{Data Collection and Preprocessing}

In the context of the Reddit platform, subreddits function as individual communities covering a variety of topics, interests, and themes. Each subreddit operates under a unique set of guidelines and regulations, designed to steer the conduct and content within that specific community. Within these individual subreddits, users can create posts and comment under the posts to participate in specific discussions.\footnote{\url{https://support.reddithelp.com/hc/en-us/categories/200073949-Reddit-101}}

To facilitate data collection and analysis of public perceptions regarding AI, we employ the Python Reddit API Wrapper (PRAW) provided by Reddit. Through the API, we are able to crawl subreddits, posts, comments, their UTC time stamp of creation and author information from Reddit. Our first step involves the formulation of a list of keywords which encompass the prevalent terminologies frequently discussed since the launch of ChatGPT. The list includes: [``AIGC'', ``ChatGPT'', ``GPT'', ``OpenAI'', ``Bard'', ``LLM'', ``large language model'', ``Midjourney'', ``diffusion model'', ``stability AI'', ``AI'', ``artificial intelligence'', ``artificial intelligence generated content'', ``dalle 2'']. Next, we conduct a comprehensive search across all subreddits, identifying those containing any of the keywords in the list. The keyword search is not case sensitive. Subsequently, we extract posts, comments, author information, and the timestamps associated with these comments from each subreddits. To narrow our focus specifically to discussions emerging after the launch of ChatGPT, we impose a temporal constraint, limiting our investigation to the period between November 30, 2022, and June 8, 2023. Bot-generated content and duplicate entries are eliminated from our dataset by identifying the repeated patterns in the comments. Upon completion of this process, we collect a total of 33,912 comments distributed across 388 subreddits, establishing the corpus for our analysis. Table~\ref{tab:summary} presents the summary statistics of the ten subreddits that have the most comments regarding AI in our dataset. When it comes to discussions surrounding AI, the subreddit ``r/singularity'' takes the lead among all other subreddits. This subreddit's unparalleled popularity can be attributed to its staggering membership count of over 955,000 individuals who actively engage in conversations revolving around the concept of technological singularity and its associated subjects.\footnote{\url{https://www.reddit.com/r/singularity/}} Before we use specific models to investigate the research questions, we further process the corpus with the Natural Language Toolkit (NLTK) library to lemmatize the corpus and remove the links, numbers, emojis, punctuation and stop words.

\begin{table}[h]
\caption {Summary statistics of top 10 subreddits with most comments.} 
\label{tab:summary} 
\centering
\begin{tabular}{l*{3}{c}r}
\hline
Subreddit              & \# comments & \# authors \\
\hline
r/singularity & 6,797 & 2,674 \\
r/technology &	2,440 &	1,864 \\
r/ArtificialInteligence &	1,761 &	846 \\
r/Futurology &	1,498 &	1,020 \\
r/artificial &	1,259 &	727 \\
r/aiwars	& 1,227 &	249 \\
r/OpenAI &	1,177 &	736 \\
r/wallstreetbets	 & 1,087 &	740 \\
r/ChatGPT	 & 853	& 593 \\
r/slatestarcodex	& 665 &	275 \\
\hline
Total	& 18,764	& 9,724 \\
\hline
\end{tabular}
\end{table}

\subsection{Topic Modeling and Cross-subreddit Analysis}
To identify the topics that characterize the discussions of AI, we leverage BERTopic~\cite{grootendorst2022bertopic}. This approach demonstrates proficiency in extracting topic representations by employing the class-based TF-IDF procedure, integrating Sentence-BERT~\cite{thakur-2020-AugSBERT} for embedding, as well as HDBSCAN~\cite{McInnes2017} for clustering within its framework. While alternative methods for topic modeling such as Latent Dirichlet Allocation (LDA)~\cite{wu2021characterizing}, Non-negative Matrix Factorization (NMF)~\cite{gan2021non}, and Top2Vec~\cite{angelov2020top2vec} are also widely used, BERTopic stands out for its exceptional efficiency in analyzing social media text data~\cite{egger2022topic}. Furthermore, BERTopic is specifically chosen for this study due to its capability in grasping and portraying context, which is a critical element in our research.

To capture more detailed insights into user attitudes towards AI, comments are particularly valuable due to their longer and more comprehensive nature compared to posts. Therefore, we focus our topic analysis exclusively on the comment corpus. The embedding model for BERTopic is {\tt all-MiniLM-L6-v2}. We identify 232 topics in total, and 16 of them are considered outlier topics. For each comment in the outlier topic, we select the most frequent topic in that comment based on topic distributions. The most frequent topic is then assigned as the topic for that comment. 

\begin{figure*}[t]
    \centering
    \includegraphics[width=0.8\textwidth]{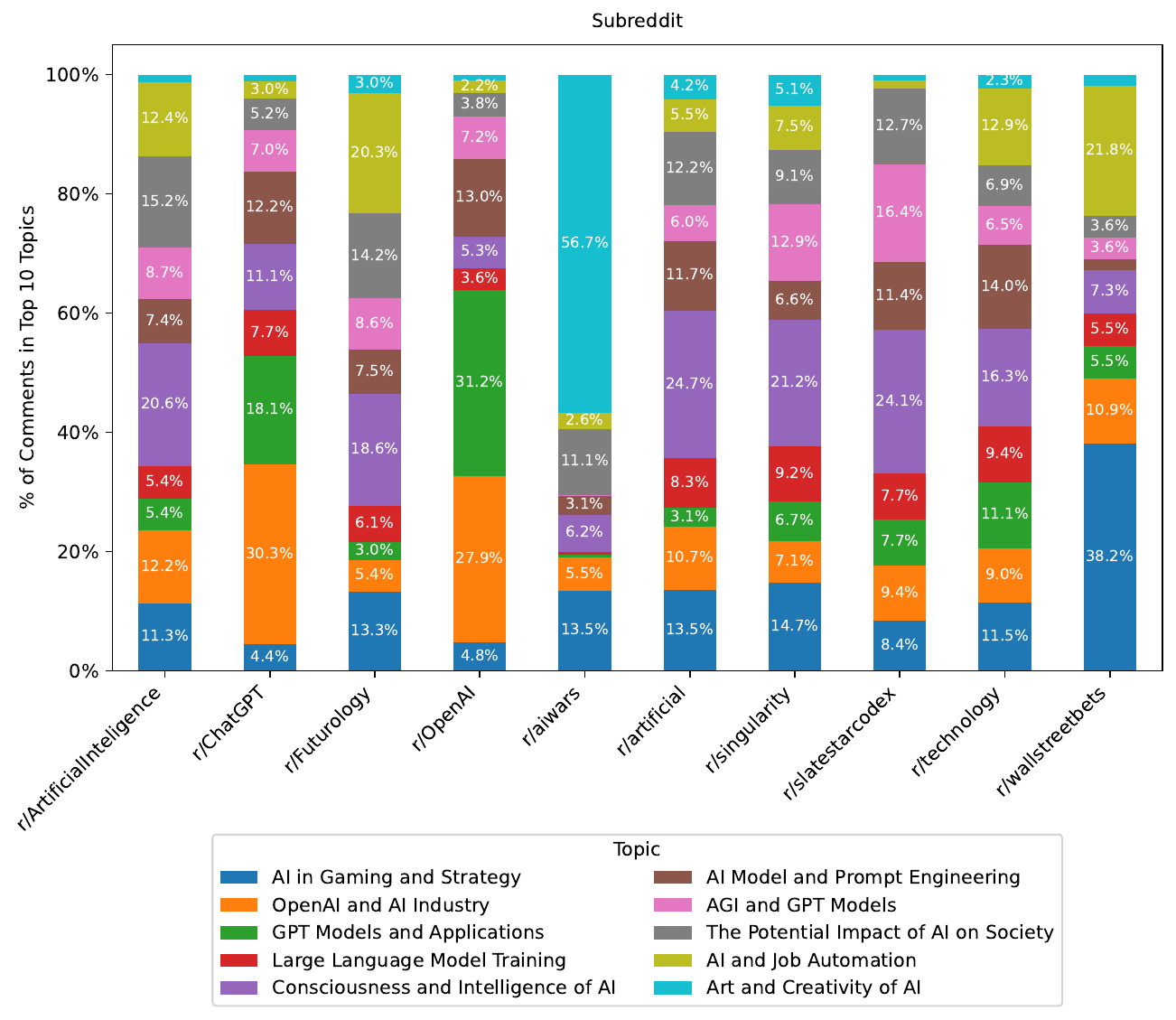}
    \caption{Topic distributions across subreddits.}
    \label{fig: topic_subreddit}
\end{figure*}

After identifying the key topics related to AI on Reddit, we further exploit the hierarchical structure of the Reddit platform, encompassing subreddits, posts, and comments to study the perceptions across different communities. Firstly, we examine the similarity and variation of topics across different subreddits, uncovering the topics that are discussed among multiple groups.  Next, we delve into the disparities between tech-centric and non-tech communities. We employ the Linguistic Inquiry and Word Count (LIWC)~\cite{tausczik2010psychological} to capture the linguistic and psychological characteristics within the comments and implement a linear regression analysis to quantify the differences between the two groups. The linear regression is specified as follows: 

\begin{equation}
LIWC \: Attribute = \alpha_0 + \alpha_1 \cdot Tech + \epsilon
\end{equation}

where $LIWC \: Attribute$ is the continuous measurement generated by LIWC. LIWC operates by categorizing words into different linguistic and psychological dimensions. It includes a comprehensive dictionary containing words that are associated with specific categories or dimensions. The software can analyze the frequency and distribution of these words in a given text and provide information about the psychological, emotional, and cognitive aspects reflected in the text~\cite{tausczik2010psychological}. In our analysis, we use the {\tt Tone}, {\tt Emotion}, {\tt Prosocial} and {\tt Conflict} merics as the outcomes. {\tt Tone} represents the degree of positive/negative tones of the corpus, while {\tt Emotion} is ``true emotion labels, as well as words that strongly imply emotions.'' {\tt Prosocial} is the category of behaviors or indicators that demonstrate assistance or empathy towards others, specifically at an interpersonal level. Lastly, {\tt Conflict},  refers to concepts that indicate or embody conflict in a broad sense~\cite{boyd2022development}.

$Tech$ is a binary variable that indicates whether the comment belongs to a tech-centric subreddit, and $\epsilon$ is the error term. Among the ten subreddits that have the most comments, we define the subreddit ``r/singularity'', ``r/technology'', ``r/Artificialintelligence'', ``r/artificial'', ``r/aiwars'', ``r/OpenAI'', ``r/ChatGPT'' as tech-centric subreddits in accordance of their names and descriptions. We assign value of 1 to the $Tech$ variable regarding the comments from these subreddits. Other subreddits, including ``r/Futurology'', ``r/wallstreetbets'' and ``r/slatestarcodex'' are regarded as non-tech subreddits and the $Tech$ variable regarding their comments is assigned 0.

\subsection{Sentiment Analysis}

\begin{figure}[t]
    \centering
    \includegraphics[width=0.5\columnwidth]{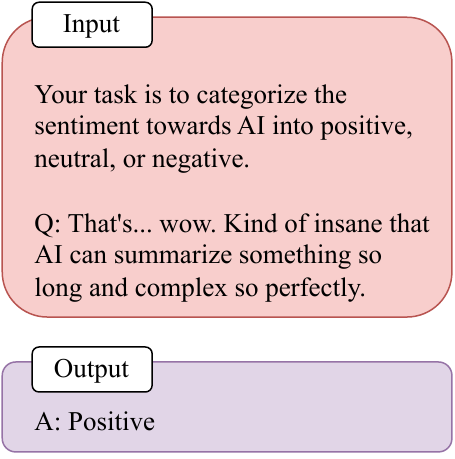}
    \caption{An example prompt for sentiment prediction.}
    \label{fig: prompt}
\end{figure}

We perform sentiment analysis to discern the attitudes and perception of each user towards AI.
\subsubsection{Modeling} ChatGPT ({\tt gpt-3.5-turbo})~\cite{ouyang2022training} is applied to our collected data and classify user comments into three sentiment categories: positive , neutral, and negative. We use zero-shot prompting and an example prompt is demonstrated in Figure~\ref{fig: prompt}. {\tt Temperature} is set at 0 to encourage the response from ChatGPT to be more focused and deterministic. Next, the final sentiment inference generated by ChatGPT is attained by employing a keyword matching approach. Upon identifying the presence of {\tt positive}, {\tt neutral} or {\tt negative} within ChatGPT's response, we assign the corresponding sentiment label to the comment as positive, neutral, or negative. Following the sentiment predictions, we proceed to analyze the individual contributions of each topic to both positive and negative sentiments. This analysis involves calculating the following ratio: $p = \frac{n_{t}}{N_s}$, where $N_s$ represents the total count of positive or negative sentiment comments, while $n_t$ denotes the count of positive or negative sentiment comments associated with a particular topic or subreddit.

\subsubsection{Performance Verification}
To evaluate the performance of ChatGPT, we randomly sample 200 comments and two researchers independently annotate these comments into the three sentiment categories. The final annotation is derived through a consensus reached between the two researchers. The Cohen's Kappa is 0.63 which suggests a substantial level of agreement~\cite{landis1977measurement}. The F1 score of ChatGPT is 0.7, indicating a commendable performance in sentiment classification on our collected dataset.

\begin{table*}[ht!]
\caption {Top 10 most commented topics identified by BERTopic.} 
\label{tab:bertopic} 
\centering
\resizebox{\linewidth}{!}{
\begin{tabular}{l*{3}{c}r}
\hline
Topic              & \# comments & Keywords \\
\hline
Consciousness and Intelligence of AI & 1,488 & consciousness, brain, conscious, human, intelligence, understand, experience, definition, complex \\
AI in Gaming and Strategy & 1,424 & game, play, player, video, chess, want, time, like, win, start \\
AI Model and Prompt Engineering & 1,150 & text, context, prompt, data, task, information, response, token, model, answer \\
OpenAI and AI Industry & 1,112 & openai, open, source, altman, api, microsoft, company, model, google, regulation \\
GPT Models and Applications & 965 & gpt, plugin, use, prompt, answer, access, ask, try, api\\
The Potential Impact of AI on Society & 893 & life, technology, potential, individual, world, lead, include, impact, society, human \\
Large Language Model Training & 771 & llm, local, data, train, agi, hallucination, parameter, different, think \\
Art and Creativity of AI & 745 & art, artist, draw, artwork, piece, artistic, ai, image, create, artist \\
AGI and GPT Models & 701 & intelligence, agi, gpt, human, general, architecture, intelligent, smart, task, asi \\
AI and Job Automation & 581 & job, replace, worker, automate, collar, work, automation, white, ai, people\\

\hline
Total	& 9,830	&  \\
\hline
    \end{tabular}}
\end{table*}

\section{Results}

\subsection{RQ1 Results}

\subsubsection{BERTopic Modeling Results}
Table~\ref{tab:bertopic} presents the ten most frequently discussed topics, which comprise 29\% of all discussions. The contribution of each keyword to its topic is listed in descending order. Although these keywords are provided by the BERTopic model, we adhere to the convention in topic modeling research~\cite{wu2021characterizing, okon2020natural} and manually assign a label to each topic based on its associated keywords. Among all the identified topics, those relating to the {\tt Consciousness and Intelligence of AI} have the highest number of comments. To provide an intuitive sense of the discussions under this prevalent topic, we present an example comment as follows:

{\it ``I am not sure if you meant to imply that is the current state of AI? If so, then that is incorrect. Humans have not developed self-aware AI programs (yet). We don't really know if that's possible to realize yet. Also, self-aware AI is pretty hard to define.''}\footnote{The example comments of all top 10 topics are listed in the Appendix.}

The prevalence of the topic suggests that the potential for AI to possess or develop human-like awareness gains substantial attention on Reddit. 

Other topics attracting significant attention include AI development and model training (\eg {\tt AI Model and Prompt Engineering}), AI in business (\eg {\tt OpenAI and AI Industry}), the creativity engendered by AI (\eg {\tt Art and Creativity of AI}), and the potential societal influence of AI (\eg {\tt AI and Job Automation}). These findings indicate that Reddit users prioritize both the \textbf{technical progression of AI and its social implications} as significant areas of interest and attention.

\subsubsection{Cross-subreddit Analysis}
To uncover whether different groups of people focus on different topics, we further conduct a cross-subreddit analysis and investigate the topic and user base difference across diverse subreddits. Figure~\ref{fig: topic_subreddit} illustrates the distribution of the top 10 topics across the ten subreddits that elicited the most comments. While {\tt Art and Creativity of AI} and {\tt AI in Gaming and Strategy} are dominant in ``r/aiwars'' and ``r/wallstreetbets'', respectively, other subreddits reveal more evenly distributed topics. Remarkably, three topics emerge as particularly prevalent, each accounting for at least 20\% of the discussions in at least two subreddits: {\tt Consciousness and Intelligence of AI}, {\tt OpenAI and AI Industry}, and {\tt AI and Job Automation}. Furthermore, while technology-centric subreddits such as ``r/ArtificialIntelligence'' and ``r/technology'' are primarily focused on AI's technical advancements, non-technology subreddits such as ``r/Futurology'' and ``r/wallstreetbets'' exhibit a higher level of interest in the social implications of AI.

\subsection{RQ2 Results}

\subsubsection{Sentiment across Topics}
\begin{figure*}[t]
    \centering
    \begin{minipage}{0.49\textwidth}
    \includegraphics[width=\columnwidth]{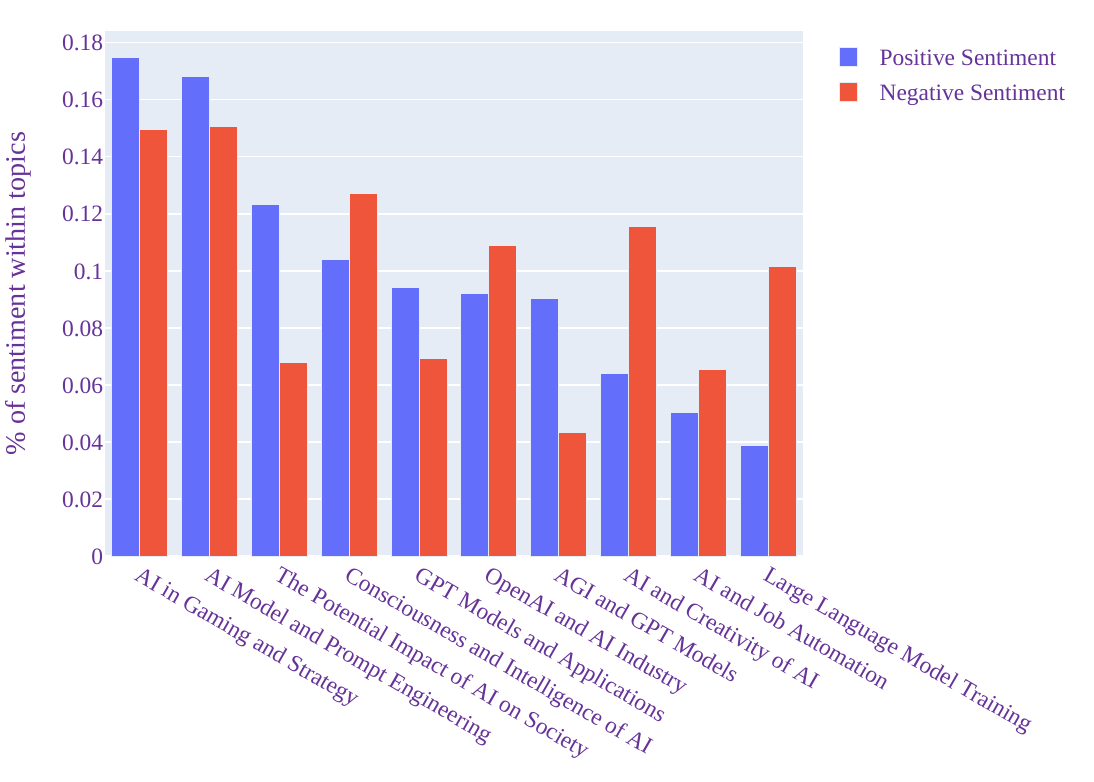}
    \caption{Sentiment distributions across topics.}
    \label{fig:sentiment_topic}
    \end{minipage}\hfill
    \begin{minipage}{0.49\textwidth}
    \centering
    \includegraphics[width=\columnwidth]{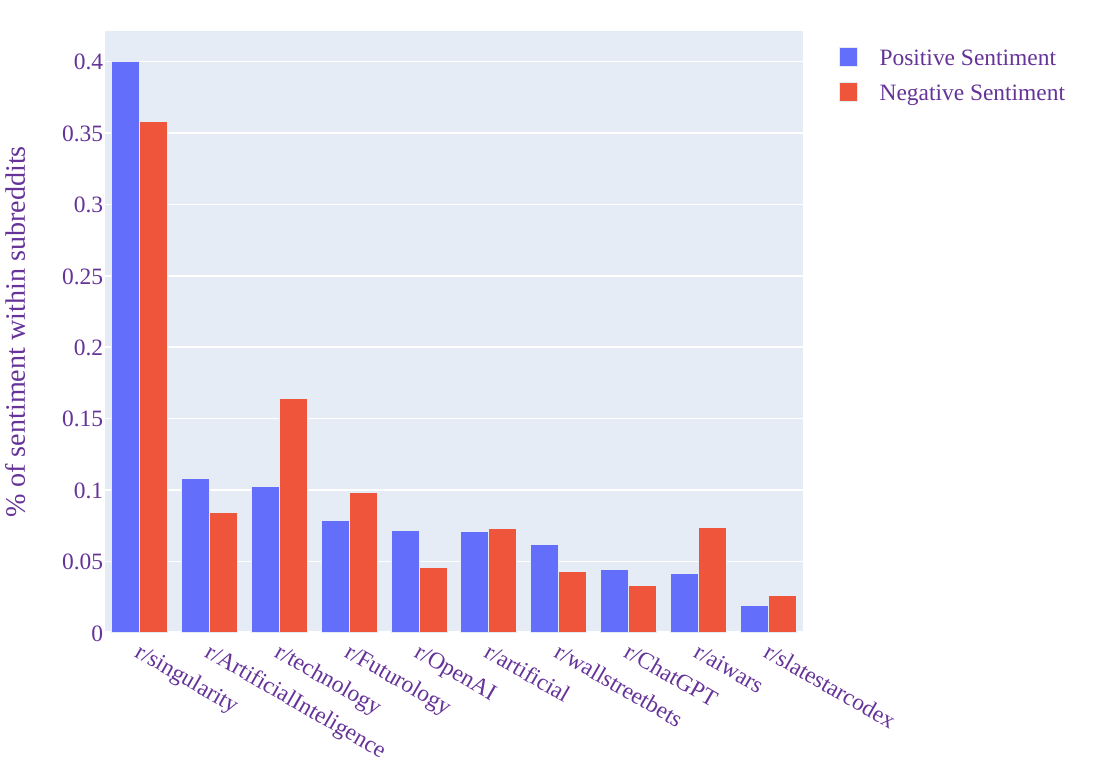}
    \caption{Sentiment distributions across subreddits.}
     \label{fig:sentiment_subreddit}
    \end{minipage}\hfill
\end{figure*}

To further investigate the sentiment differences in terms of topics, we compute the percentages of positive and negative comments of each topic.  Figure~\ref{fig:sentiment_topic} displays the percentage of positive/negative comments. The topics of {\tt AI in Gaming and Strategy}, {\tt AI model and Prompt Engineering}, {\tt GPT Models and Applications}, {\tt The Potential Impact of AI on Society}, and {\tt AGI and GPT Models} exhibit a higher percentage of positive sentiment compared to negative sentiment. In contrast, the topics of {\tt Consciousness and Intelligence of AI}, {\tt OpenAI and AI Industry}, {\tt Large Language Model Training}, {\tt AI and Creativity of AI}, and {\tt AI and Job Automation} exhibit higher percentages of negative sentiment. Based on the data presented in Table~\ref{tab:bertopic}, it is evident that the topics with higher percentages of positive sentiment primarily emphasize the benefits and conveniences offered by AI. Conversely, the topics exhibiting higher percentages of negative sentiment tend to focus on the drawbacks and future development of AI.

The aforementioned findings indicate a positive reception of AI applications and a general comprehension of the potential uses of AI models. The public holds the belief that AI can contribute to the betterment of society, particularly when employed as an assistant in decision-making processes, such as gaming and education. To illustrate, here is an example comment reflecting a positive sentiment from the topic {\tt The Potential Impact of AI on Society}:

{\it ``For what it’s worth, this method of research has existed for a long time. It’s called High Throughput Testing. It’s basically a `throw everything at the wall and see what sticks' approach. I think using AI to test drugs we wouldn’t have otherwise thought of, to test a higher quantity of drugs, and to analyze the efficacy of those drugs is overall a great idea. Of course it will always require humans to verify the results and make final clinical decisions.''}

In this particular example, AI is expected to assist in conducting drug tests due to its impressive capabilities. However, given the current limitations of the AI system, human verification remains necessary. Despite this limitation, the overall sentiment towards AI remains positive.

The negative comments highlight a lack of trust in current AI technology and apprehension regarding the future trajectory of AI development. Particularly within topics like {\tt Consciousness and Intelligence of AI} and {\tt AI and Creativity of AI}, the public expresses concerns regarding potential issues arising from AI. Keywords from Table~\ref{tab:bertopic} reveal problems such as regulation, hallucination, and job replacement, while security and privacy are recurring themes in these negative comments. Here is an example comment that reflects the sentiment of mistrust towards AI:

{\it ``Yes, but there is a difference between understanding and relaying information. GPT 4 can relay information well, but that doesn’t mean it actually understands what it is doing. It’s just tossing around words in an organized manner based on the the prompt that you give it. So basically, it isn’t making it’s own thoughts, it’s just re-engineering words and sequences to make it seem like it’s making new thoughts. Until we learn about the actual nature of consciousness (if there even is one) A.I. is just another marketing buzzword.''}

In this specific example, the sentiment expressed is a lack of trust in AI, as people question whether AI (specifically GPT-4) truly embodies real intelligence. Other studies have arrived at similar conclusions. For instance, Beets \etal~\cite{info:doi/10.2196/40337} highlighted that in the healthcare domain, individuals reap the benefits of AI advancements, but they exercise caution when AI is involved in making critical personal health decisions. Additionally, Zhang and Dafoe~\cite{zhang2019artificial} proposed that the public supports the development of AI due to its promising potential, yet they also express the belief that AI should be subject to careful management.

\subsubsection{Sentiment across Subreddits}

Figure~\ref{fig:sentiment_subreddit} shows the percentage of positive/negative comments of each subreddit. It is noticable that among the analyzed subreddits, namely ``r/technology'', ``r/Futurology'', ``r/aiwars'', ``r/artificial'', and ``r/slatestarcodex'', which have relatively uniform topic distributions, there is a higher proportion of negative sentiment compared to positive sentiment. The prevalent discussion of AI within these subreddits indicates that Reddit users thoroughly engage in diverse AI-related topics, expressing their viewpoints on the current limitations of AI, including concepts like ``misinformation'' and ``stochastic parrot''. Additionally, the societal implications of AI are of significant concern to the public. This is exemplified by the vibrant online community of ``r/aiwars''. As for the topic of {\tt Art and Creativity of AI}, conversations primarily revolve around AI-generated text and images, which have emerged as dominant subjects of interest. However, the surge in AI creativity has also given rise to derivative issues, such as unemployment and copyright concerns, stimulating thoughtful deliberations among community members.

Conversely, ``r/singularity'', ``r/ArtificialIntelligence'', ``r/OpenAI'', ``r/wallstreetbets'', and ``r/ChatGPT'' demonstrate the opposite pattern. According to Figure~\ref{fig: topic_subreddit}, ``r/ChatGPT'' accounts for approximately 46\% of comments associated with the topics characterized by positive sentiment. People generally hold a positive appreciation for AI, recognizing its capabilities and convenience. A prime example is ChatGPT, which serves as a valuable writing tool, assisting users in generating high-quality text. As a result, business professionals can foresee the potential of AI enhancing their productivity through its intelligent capabilities. This positive sentiment reflects the recognition of AI's strengths and its potential to positively impact various aspects of human endeavors. These findings suggest that users perceive and engage with various topics in distinct ways, leading to differing sentiments across subreddits.

\subsubsection{Tech-centric vs. Non-tech Subreddits Analysis}

\begin {table*}[ht!]
\begin{center}
\caption {Regression analysis across tech and non-tech groups}
\label{tab:regression_tech_non_tech}
{
\def\sym#1{\ifmmode^{#1}\else\(^{#1}\)\fi}
\begin{tabular}{l*{6}{c}}
\hline\hline
            &\multicolumn{1}{c}{(1)}&\multicolumn{1}{c}{(2)}&\multicolumn{1}{c}{(3)}&\multicolumn{1}{c}{(4)}&\multicolumn{1}{c}{(5)}&\multicolumn{1}{c}{(6)}\\
            &\multicolumn{1}{c}{Positive tone}&\multicolumn{1}{c}{Negative tone}&\multicolumn{1}{c}{Positive emotion}&\multicolumn{1}{c}{Negative emotion}&\multicolumn{1}{c}{Prosocial}&\multicolumn{1}{c}{Conflict}\\
\hline
\\
$Tech$        &       0.173\sym{**} &     -0.0160         &       0.0562\sym{*} &       0.102\sym{***}&      0.0632\sym{*} &      0.0153         \\
            &    (0.0630)         &    (0.0519)         &    (0.0282)         &    (0.0297)         &    (0.0287)         &    (0.0205)         \\
            \\
\multicolumn{2}{c}{Reference Group: Non-tech subreddits} \\

\hline 

\(N\)       &       18764         &       18764         &       18764         &       18764         &       18764         &       18764         \\
\hline\hline
\multicolumn{7}{l}{\footnotesize \begin{minipage}{0.8\linewidth} \smallskip \textbf{Note:} This table presents the estimation coefficients of the regressions. Standard errors of each coefficient are in parentheses. The p-values indicating significance at the 90\%, 95\%, and 99\% confidence levels have been adjusted using the Bonferroni correction.   \end{minipage}}\\
\multicolumn{7}{l}{\footnotesize \sym{*} \(p<0.1\), \sym{**} \(p<0.05\), \sym{***} \(p<0.01\)}\\
\end{tabular}
}
\end{center}
\end{table*}

Another question that we aim to explore is whether perceptions of AI differ between tech-centric and non-tech communities. To evaluate the emotional and social attributes of comments, we leverage LIWC (Linguistic Inquiry and Word Count), which is a software for analyzing word use and can be used to study a single individual and groups of people, utilizing its built-in dimensions related to psychological and social processes. More specifically, we examine six key dimensions: \textit{positive tone}, \textit{negative tone}, \textit{positive emotion}, \textit{negative emotion}, \textit{social interactions}, and \textit{interpersonal conflict}. It is worth noting that tones distinct with emotions in the way that they only captures sentiment, while emotions are restricted to estimqte the words that strongly imply emotions~\cite{boyd2022development}. Table~\ref{tab:regression_tech_non_tech} shows the regression results regarding each dimension between tech-centric and non-tech subreddits.

The outcomes displayed in columns (1) and (2) in Table~\ref{tab:regression_tech_non_tech} shows the discrepancies in the use of positive and negative tones in comments. Based on these results, it can be deduced that the positive tone scores of the comments from the tech-centric communities, on average, are 0.173 higher than their non-tech counterparts. However, there is no similar disparity found in the regression results regarding the negative tone. This suggests that tech-centric communities exhibit a greater level of optimism in tones regarding AI advancements compared to non-tech communities. 

In terms of emotional expression, as depicted in columns (3) and (4), tech-centric communities reveal higher scores in both positive and negative emotional expressions. These findings suggest that \textbf{tech-centric communities are more polarized in their sentiments compared to non-tech-centric communities.} This disparity can be attributed to the familiarity with technology and the resulting tendency for more distinct and definitive expressions within the tech-centric communities.

Furthermore, the tech-centric communities demonstrate a higher score in prosocial behaviors, suggesting that these communities exhibit a greater inclination towards expressing signals of ``helping or caring about others'' in their discussions about AI, This observation is based on the LIWC psychometric measurements~\cite{boyd2022development}. On one hand, the prosocial expressions could stem from the inherent collaborative spirit found within the tech-centric communities, such as the open-source software culture. On the other hand, they could be a result of concerns regarding the ethical implications and potential impact on societal well-being stemming from AI advancements.

\section{Discussion and Conclusion}
In this study, we delve into understanding the public perception of artificial intelligence, utilizing a dataset of 33,912 comments from 388 unique subreddits, spanning from the launch of ChatGPT on November 30, 2022, to June 8, 2023. Employing BERTopic, we uncover a wide range of diverse topics discussed on Reddit, surpassing the findings of existing literature on public perception of AI~\cite{miyazaki2023public,leiter2023chatgpt,tlili2023if,sullivan2023chatgpt}. The most frequent topics include the discussions about the consciousness and intelligence of AI, AI development and model training, AI in business, the creativity engendered by AI, and the potential societal influence. 

The results from our sentiment analysis reveal nuanced variations in sentiment across different subreddits and topics. Overall, the public tends to perceive AI as a beneficial force that can contribute to societal improvement, particularly when used as an assistant in decision-making processes like gaming and education. However, negative comments highlight a lack of trust in current AI technologies and apprehension about the future trajectory of AI development, \textit{aligning} with the findings of Leiter~\etal~\cite{leiter2023chatgpt}. Furthermore, LIWC is employed to examine the more fine-grained differences in sentiment between the tech-centric and non-tech communities. We find that tech-centric communities exhibit higher polarization in their sentiments compared to non-tech-centric communities. While our analysis uncovers differences across subreddits, which serve as proxies for distinct social groups, further investigations could explore the underlying factors contributing to these differences.

In conclusion, this study on understanding public perception of AI has shed light on the multifaceted landscape of opinions, attitudes, and concerns surrounding artificial intelligence. Through various research methodologies, we have discovered the prevalent topics, sentiments, and thematic variations across different communities. The findings emphasize the importance of considering public opinion in shaping AI policies, addressing ethical considerations, driving user acceptance, promoting education and awareness, and guiding the design and development of AI technologies. This comprehensive understanding of public perception serves as a valuable foundation for fostering responsible and beneficial AI innovations that align with societal expectations and values. By bridging the gap between AI development and public sentiment, we can work towards building a future where AI technologies are embraced, trusted, and utilized in a manner that positively impacts individuals and society as a whole.

\bibliographystyle{ieee_fullname}
\bibliography{egbib}

\appendix
\subsection{Example Comments of the Top 10 Most Prevalent Topics}

\small \textbf{Consciousness and Intelligence of AI: } ``I am not sure if you meant to imply that is the current state of AI? If so, then that is incorrect. Humans have not developed self-aware AI programs (yet). We don't really know if that's possible to realize yet. Also, self-aware AI is pretty hard to define.'' \\

\small \textbf{AI in Gaming and Strategy: } ``The purpose of human brains is not to play chess. The game of chess is just one of uncountable activities they can learn to do, because of their extreme flexibility. The complexity of analysis that every conscious brain performs every second outperforms any AI to incredible extents and it's going to stay that way for a long time. Of course a specialized AI may be better at specialized tasks, like playing a certain game. But it's still very limited machine. Machines are often better at their specialized task, than humans, but a single machine won't be able to do a fraction of activities, than a human is able to do. AI trained to play chess is just that, machine to play chess - it want be able to consciously adapt to any other task.'' \\

\small \textbf{AI Model and Prompt Engineering: }``It's not really in a usable state currently. Basic prompt to ChatGPT or GPT-4 usually gives better results as compared to autogpt. So it's not worth the hassle to read so many prompts, give it human feedbacks and also spend money when you can just get better output for free in a much easier way. However, this experiment does have a potential to become useful in future. One of the ways this could be done is by using adding more agents where each agent is specialized in a single task instead of something general like ChatGPT. Also, I heard that someone is working on re-implementation of AutoGPT as a python package which is a good idea in my opinion as it would allow AutoGPT to be used in actual projects.'' \\

\small \textbf{OpenAI and AI Industry: }``OpenAI was always going to do that, and the Google memo is wishful thinking since Google search profits will shrink if everyone has a quality free LLM. It could make a lot of sense for some other corporations to co-operate on a free LLM, but that's only likely if OpenAI is asleep at the wheel when it comes to undercutting competition. OA has a big lead on quality and the funding to drive prices low. Even if OA R\&D falters the new winner will have the same plan. Outside of LLMs it's quite not so dire for competitive local AI, but even there we're still totally dependent on charity for expensive base models.'' \\

\small \textbf{GPT Models and Applications: } ``Gotcha, I just gave it a try.  It got farther than normal restricted mode, but it still stopped once the story was going to get good, LOL.  Hard to say if a pinned prompt would help or not.  I also feel like ChatGPT can maybe build story elements better for now... I just get the feeling GPT-4 is too concise, but well, once I get API access it'd be fun to play around with for sure.  Anyway, I will see when I have time to work on more of those additions for the app. I'll ping you when I have an update to push. For now, I should get back to work, lol.'' \\

\small \textbf{The Potential Impact of AI on Society:} ``AI means the end of reliably documented events. There's just no way around it, sadly. There is no longer any objective reason for anyone to believe anything anybody tells them from now on. The "post-truth" era is real and it's here. What that really means is we can no longer afford to assume any organized system of authority has our best interests in mind.'' \\

\small \textbf{Large Language Model Training:} ``having 95\% ChatGPT performance is a strong claim to have and easy to disspell. Also, I have strong doubts about the propietary model that's a delta weight  from LLaMA. I think the licence from LLaMA makes that illegal but I'm not that much in the know. Logic problems are a metric for finding how useful these models are. If you want to test your model against others, instead of making wild claims submit it to huggingface's LLM leaderboard'' \\

\small \textbf{Art and Creativity of AI:} ``The difference with using stock photos and such is you either already had the free rights to use them or payed to use them. Stock assets are there to be used. AI generated media is based on training sets composed of stolen and uncredited art, many times without an original artists consent. While its amazing technology its got a very shaky ethically questionable foundation. The very least that could have been done is noting the piece as AI assisted / generated'' \\

\small \textbf{AGI and GPT Models:} ``I wonder if anyone knows and can say. When they would create lets say GPT-5 would it be a completely new system from the ground up? Or would it be a sort of upgrade building on the previous version? I mean obviously no one knows. I guess i am just asking if it is common to `upgrade' systems in this way, or are usually just new models made? I know they have talked about upgrading GPT like in 0.1 increments etc. I guess it depends and if they also make some changes in architecture.'' \\

\small \textbf{AI and Job Automation:} ``the positions in the industry will already be consolidated and companies will be operating with a skeleton crew. This was said with respect to every disruptive technology ever. The reality is that people will move into roles where they manage the workload of AI rather than doing the work themselves. Employment isn't some magic thing that fell from the sky. It's the thing we invent and re-invent constantly in order to demonstrate value to others. If all of the jobs everywhere go away, we'll just invent more of them because it's what we do. Suggesting otherwise is like saying that once AI shows up, children won't play in the yard because AI can do that.''

\end{document}